# Broad Band Single Germanium Nanowire Photodetectors with Surface Oxide Controlled High Optical Gain

Shaili Sett[1], Ankita Ghatak[1], Deepak Sharma[2], G. V. Pavan Kumar[2] and A. K. Raychaudhuri[1*]

[1]Department of Condensed Matter Physics and Material Sciences, S.N.Bose National Centre for Basic Sciences, JD Block, Sector –III, Kolkata 700106, India.

[2]Division of Physics, Indian Institute of Science Education and Research, Pune 411008, India.

*Email: arup@bose.res.in

**ABSTRACT:** We have investigated photoconductive properties of single Germanium Nanowires (NWs) of diameter < 100 nm in the spectral range of 300-1100 nm and in the broadband Near Infra-red spectrum showing peak Responsivity ($\mathcal{R}$) ~$10^7$A/W at a minimal bias of 2V. The NWs were grown by Vapor-Liquid-Solid method using Au nanoparticle as catalyst. In this report we discuss the likely origin of the ultra large $\mathcal{R}$ that may arise from a combination of various physical effects which are (a) Ge/$GeO_x$ interface states which act as "scavengers" of electrons from the photo-generated pairs, leaving the holes free to reach the electrodes, (b) Schottky barrier (~0.2 -0.3eV) at the metal/NW interface which gets lowered substantially due to carrier diffusion in contact region and (c) photodetector length being small (~few μm), negligible loss of photogenerated carriers due to recombination at defect sites. We have observed from power dependence of the optical gain that the gain is controlled by trap states. We find that the surface of the nanowire has presence of a thin layer of $GeO_x$ (as evidenced from HRTEM study) which provide interface states. It is observed that these state play a crucial role to provide a radial field for separation of photo-generated electron–hole pair which in turn leads to very high effective photoconductive gain that reaches a value > $10^7$ at an illumination intensity of 10 μW/cm$^2$.

## 1. INTRODUCTION

Achieving ultra-high response in a photodetector is of great interest in a variety of photonic applications where small radiation powers need be detected. The applications of photodetectors widen if the detector can work in a broad spectral range that spans UV-Visible and Near Infrared (NIR).[1] Semiconductor photodetectors are most widely used for this particular application.[2,3] However, most commercially available detectors made from bulk semiconductors, show responsivity ($\mathcal{R}$), (defined as the ratio of photocurrent ($I_{Ph}$) to incident light on the NW ($P$)) < 1 A/W. Recently it has been established that photodetectors made from single semiconductor nanowires (NWs) with diameters ≤ 100nm can reach extremely large photo-response with $\mathcal{R} > 10^4$A/W.[4-6] For example, Si[4], ZnO[5], GaN[6], InAs[7] and even for molecular material like Cu:TCNQ[8], provided the distance between electrodes is around a μm or less. The photoconductive gain ($G_{Pc}$) (defined as the ratio of ($I_{Ph}$) to the incident number photons of frequency ($\nu$) in unit time) in these NW devices reaches a very high value ~$10^6$ -$10^8$, which often reduces at higher illumination intensity.[5] This arises mainly due to the role played by traps in the surface depletion layer which provide a radial field that separates the photo-generated carriers,[5] thus mitigating carrier recombination and leading to a very high gain.

Other than ultrahigh responsivity, another desirable aspect in a photodetector is the wide band width of detection extending from UV to NIR that opens up its applications. It is envisaged that if single nanowire broad band detectors can be made they will pave the way for integrating such wires into high performance detector arrays and may even act as detectors for on-chip optical communication systems. Among the single NW photodetectors that show very high responsivity only InAs[7] and Si[4] have been tested for broad band application in the wavelength range 300 nm to 1100 nm. InAs NW with diameter in the range 30-75 nm grown by MOCVD and field effect mobil-ity (μ) ~$10^4$ $cm^2$/V.sec, with illumination intensity of 1.4 mW/$cm^2$ and at a bias of 15V, shows a value of $\mathcal{R}$ = 4.4 x $10^3$ A/W.[7] In a previous investigation from our group[4] it was observed that in a single Si NW detector, $\mathcal{R}$>1.5x $10^4$ A/W is achievable in a 80nm diameter nanowire in the spectral range 450-1000nm with a small bias of only 0.3V with illumination intensity of few mW/$cm^2$.

In a photoconductor, the responsivity $\mathcal{R}$ is directly proportional to the carrier mobility. In this context, investigation of photoconduction in Ge NW is very important since it is a material with larger $\mu$ compared to Si and interestingly, its band gap being smaller, it is expected to perform at longer wavelengths extending well into NIR region of the spectrum. It should be noted that Ge being an elemental semiconductor can be grown with more ease compared to the difficulty in growth of complex semiconductor NWs like InAs or InGaAs where stoichiometry control is crucial. Inspite of these superior properties, Ge NW has not been investigated till date as a photodetector over a spectral range that extends to NIR region. However, crystalline bulk Ge has been studied extensively in the VIS-NIR region (upto wavelength 1800nm).[9,10] Ge photodetectors based on single crystalline Ge are available commercially. They work in the VIS-NIR region and have peak responsivity with R ~0.25 A/W at 800 nm.[11] Previous reports of photodetectors made from Ge NWs, single[12-14] as well as ensemble[15], have been investigated only around the visible range of the spectrum with wavelength < 532nm. The highest responsivity reported is around 2x$10^2$ A/W at 2V bias.[14] An investigation of absorbance of single Ge NW (using photo current as a detection tool) showed that the photo absorption in Ge NW depends on its diameter and differs qualitatively from bulk[16] as it shows a peak around 600-800nm (depending on the diameter) and drops for wavelength > 1000nm. The change in spectral dependence was related to leaky resonance mode in NWs with lengths comparable to light wavelengths.

In this report we show that photodetectors made from a single strand of Ge NW (diameter ~ 30 nm) can reach $\mathcal{R}$ > $10^7$ A/W over an extensive spectral range 350nm -1100nm with a bias of only 2V. This value of $\mathcal{R}$ is the highest observed in single NW photodetectors till date over a broad spectral range. The peak value of $\mathcal{R}$ ~$3x10^7$A/W is observed at a wavelength of 850 nm and low illumination intensity of 10μW/$cm^2$. The detector has a finite gain down to an illumination intensity of 1μW/$cm^2$ which corresponds to detection of power of ~ $10^{-15}$ W (calculated from area of the device). It is also found that the response though reduces at longer wavelength, remains substantially high at wavelengths> 1100 nm.

One of the factors that contribute to the very large optical gain is the contribution of surface/interface states. We observe from electron microscopy data that there exists a layer of $GeO_x$ on the surface of the Ge NW. It is proposed that in interfaces like Ge/$GeO_x$ existence of interface states can lead to an effective radial field that acts to separate out the electron-hole pair created by the optical illumination.[5] As illumination intensity increases, there is enhanced pair generation rate and the states at surface/interface get saturated beyond a critical illumination intensity, which then cease to act as an efficient agent for separation of the electron-hole pairs. This leads to reduction of gain as the illumination intensity increases beyond a critical value.

## 2. EXPERIMENTAL SECTION

Ge NWs were grown by vapor phase method in a dual zone furnace using gold (Au) nanoparticle (NP) as catalyst dispersed on a Silicon substrate. The Au NPs were made by dewetting a thin film of Au grown under ultra high vacuum conditions by electron beam evaporation and subsequent vacuum annealing. Growth of NWs occurs by Vapor-Liquid-Solid mechanism with the aid of Au NPs. The details are given in a previous publication from the group.[17] The Ge NWs were dis-persed on a $SiO_2$ (300nm)/Si substrate from an ethanol suspension and connected to pre-grown contact pads (Cr(5nm)/Au(60nm) made by photolithography) by a combination of Electron Beam lithography and Focused electron beam deposited Pt. Transmission electron microscopy (TEM) of 120kV (Tecnai G 2,TF-20, ST) was employed to study the structure and phase locally on a single Ge nanowire. Selected area electron diffraction (SAED) patterns using high-resolution transmission electron microscopy were carried out on a single nanowire to see the crystal symmetry of the grown nanowires. The TEM-based Gatan parallel detection electron energy loss spectrometer (EELS) was used study the chemical binding states of Ge NWs. Electrical measurements were taken by a source-meter. The photoconductivity data were taken as $(I - V)$ curves in dark as well as illumination of different intensity. The spectral response$(I - \lambda)$ was measured by measuring the device current with a given bias with illumination at different wavelengths. A standard Xenon light source was used with a monochromator for illumination in the range 300-1100 nm. The system in the wavelength range upto 1100 nm was calibrated using a Si detector with NIST traceable calibration. At longer wavelengths we used broad band source (range 300-2600nm with peak at 1000nm, intensity at 1800nm is ½ of that at 1000nm) with a fiber-coupled output and a band stop filter that blocks radiation with wavelength < 980nm.

## 3. RESULTS AND DISCUSSION

### 3.1 Structural characterization

High Resolution Transmission Electron Microscope (HRTEM) was used for structural characterization, with Selected Area Electron Diffraction (SAED) pattern showing highly oriented <111>planes (figure 1a) with cubic structure. Inset of figure 1a shows a TEM image of a typical Ge NW of diameter 45 nm. The image also shows a lower contrast region marked with black line on both sides of the NW, which may correspond to a native oxide layer which has been discussed later in section 3.5. The High resolution TEM (HRTEM) image in figure 1b, shows parallel d-planes with spacing 0.326 nm corresponding to <111> direction. The HRTEM image also shows an amorphous oxide layer ~ 5-6 nm (marked in black line in fig. 1b), where no clear fringes are present. A proper measure of the thickness of the oxide layer has been obtained from the HAADF image (see Supplementary data, Figure. S1) which gives a clear contrast difference (also shown in Fig. 7c). Thus, the Ge NWs used to fabricate photodetectors are single crystalline and highly oriented.

### 3.2 Photodetector characteristics

We have studied eight samples of photodetectors made from Ge NWs with diameters ranging from 30-90nm. However, in this paper we will discuss results on two wires with diameter 30 nm and 65 nm (see table 1 for summary of samples measured and experimentally obtained parameters). The results in other wires are qualitatively similar and the responsivity typically lies within the range of the $10^5$-$10^7$A/W. The length in the table refers to the distance between the two electrodes (referred to as $L$). The contact length may also partly contribute to the total length. However, the contact length being small, the extent of uncertainty is small.

The NWs grown from vapour phase are not intentionally doped. However, it has been known from previous studies[18-20] of growth and electrical measurements that the NWs are p-type in nature. During growth, the NWs are prone to catalyst doping (electron acceptor levels) that leads to the p-type nature. We have studied the following parameters: (a) the device current $I$ as a function of bias at a fixed wavelength by studying the $(I - V)$ curve with different illumination intensities $(\mathcal{J})$, (b) spectral dependence of photocurrent (PC) in the spectral range 300-1100 nm for a fixed bias. The PC is defined as $I_{Ph} \equiv (I - I_{dark})$ for a given bias, illumination intensity and wavelength, $I$ being the device current with illumination and $I_{dark}$the device current in dark at same bias voltage. The power incident on the nanowire $P$ for a particular intensity has been calculated from the illumination intensity $\mathcal{I}$(measured in W/cm$^2$) using the relation, $P = \mathcal{I} \times \mathcal{A}$, where $\mathcal{A}$ is the collection area of the nanowire defined as $\mathcal{A} = \pi L d$. The power absorbed by the nanowire is less than the incident power$P$.[21] The approximate fraction is estimated as 50% in absence of proper data. It is noted that $P$ is an upper limit of the absorbed power, thus underestimating $\mathcal{R}$.

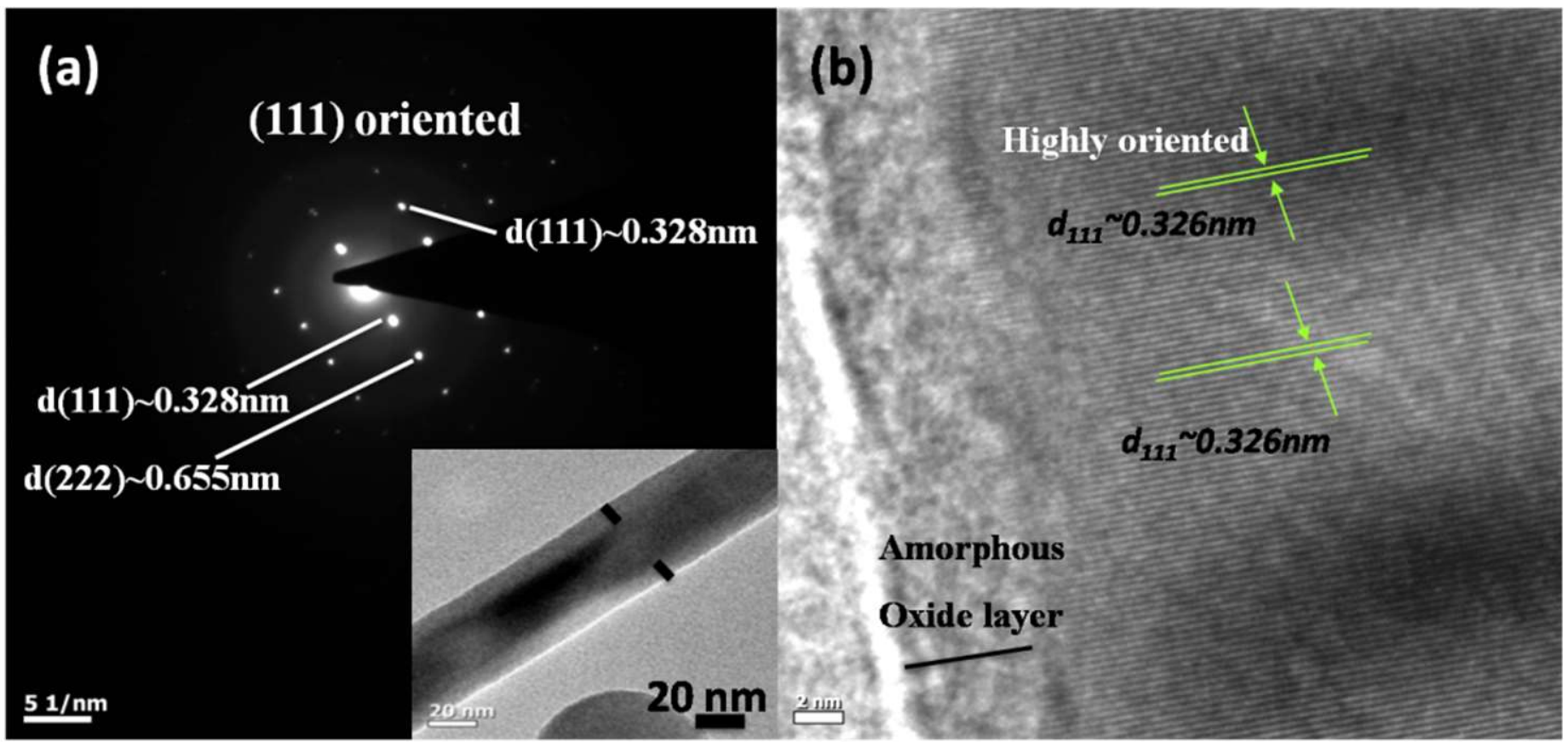

Figure 1. (a) SAED pattern of a Ge NW showing highly oriented planes along (111) direction and the inset showing a TEM image of the same, with diameter ~45 nm. The sides of the NW have been marked in black, which appear at a different contrast as compared to the core of the NW and show likely existence of oxide layer (b) HRTEM image of a Ge NW with distance between parallel d-planes~ 0.326 nm, and amorphous oxide layer, where no clear fringes are present.

The following device parameters obtained from measurements are photo-current gain ($G_{Pc}$), Responsivity ($\mathcal{R} = I_{Ph}/P$) and Specific Detectivity ($D^*$). $G_{Pc}$ at a given wavelength or frequency $\nu$ is defined as the ratio of the photo-generated carriers per unit time ($I_{Ph}$) to the incident number photons of frequency ($\nu$) in unit time ($P/h\nu$) and is given by $G_{Pc} = h\nu I_{Ph}/eP$. The Specific Detectivity is defined as, $D^* = \sqrt{\mathcal{A}\Delta f}/NEP$, where $\mathcal{A}$ is the area of the NW exposed to incident radiation, $\Delta f$ is the signal bandwidth and $NEP$ is the power at which the signal to noise ratio is unity.

(Note: The NEP includes shot noise, flicker noise and thermal noise. In most calculations of NEP, only the shot noise is calculated to determine the NEP. This is not the true value of the NEP, which should be experimentally determined as the power at which the photocurrent to dark ratio is unity.) All the relevant parameters are summarized in Table 1.

**Table 1. Ge NW photodetector device parameters measured at wavelength 650nm and illumination intensity 1000μW/cm².**

| Nanowire Name | Diameter (d) (nm) | Length ($L$) (μm) between electrodes | Resistivity in dark ($\rho o$) (Ω-cm) | Responsivity ($\mathcal{R}$)(A/W)at bias voltage in brackets | Detectivity($D$) (cmHz$^{1/2}$/W) | Photo-generated carrier ratio ($\frac{\Delta n}{n_o}$) | Gain ($G_{PC}$) | Lowering of Schottky barrier height under illumination of maximum intensity ($\|\Delta\phi\|$) |
|---|---|---|---|---|---|---|---|---|
| N1 | 65 | 3.2 | 6.5x10$^{-2}$ | 1x10$^{5}$ (1V) | 5.5x10$^{11}$ | 6.2 | 4x10$^{5}$ | 19 meV |
| A1 | 30 | 1.2 | 4.9x10$^{-2}$ | 3x10$^{6}$ (1 V)<br>2.8x10$^{7}$ (2 V) | 4.2x10$^{12}$<br>1.5x10$^{13}$ | 12.2 | 1.5x10$^{6}$<br>2.1x10$^{7}$ | 61 meV |

In Figure 2a we show a typical example of $(I - V)$ curves taken in dark and with illumination of 650nm with an intensity of 100μW/cm² on NW A1 (diameter 30 nm). At a bias of 2V the current is around 1μA. In Figure 2(b) and (c) we show a contour plot of device currents as a function of bias with different levels of illumination intensities in NW N1 and A1 respectively. The colour code gives the device current plot data with same scales of illumination and upto bias of 2V. Figure 2d shows Gain in the NWs as a function of bias voltage. The I-t curves of the photodetectors at a fixed bias are given in Supplementary data, Figure S2.

The $(I - V)$ curves of NW photodetectors can be modelled as an MSM device with back to back Schottky diodes and thermionic emission (TE) as the dominant mechanism of carrier injection across the MS interface.[22] This allows us to find two important device parameters, namely, the Schottky Barriers, $\phi_1$ and $\phi_2$at the MS interfaces and the value of nanowire resistance $R_{NW}$ which lies in between the two MS contacts. We have analyzed the $(I - V)$ curves of both devices using the modified Richardson –Dushmann equation[22] given by:

$$I = I_0\left(\exp\left(\frac{qV'}{\eta kT}\right)-1\right)\frac{\exp\left(\frac{-q(\phi_1+\phi_2)}{kT}\right)}{\exp\left(\frac{-q\phi_2}{kT}\right)+\exp\left(\frac{-q\phi_1}{kT}\right)\exp\left(\frac{qV'}{\eta kT}\right)}, \quad (1)$$

where, $I_0$ is contribution due to thermionic emission at the contacts given by $AA_RT^2$, where $A$ is the area of contact and $A_R$ is the Richardson constant, $V' = V - IR_{NW}$ and $\eta$ is the ideality factor of the diode. The analysis of the $(I - V)$ curves (fit to the data are shown as solid lines in Figure 2a) allow us to determine the evolution of $R_{NW}$ with illumination intensity. This helps to determine the extra carrier density $(\Delta n)$ generated by illumination. The analysis also gives us a measure of the lowering of the barrier height on illumination. This is an important factor that adds on to the photocurrent and enhances it. $R_o$ is designated as the value of resistance of NW in dark and from it we obtain the resistivities of the NWs. It can be seen from table 1, that the resistivity $\rho_o$of NW N1 $6.5 \times 10^{-2}\Omega$cm and A1 is $4.9 \times 10^{-2}\Omega$cm. A comparison of the resistivities of the NWs with bulk[20] will give an approximate carrier concentrations $(n_o)$ in the NWs in dark and at ~ 1000 μW/cm$^2$ respectively.

The enhancement in conduction of the NW ($\Delta g_{NW}$) under illumination has been derived from the relation, $\Delta g_{NW} = (R_{NW}^{-1} - R_0^{-1})$. Both $R_{NW}$ and $R_o$ have been obtained from experiment. The ratio $\delta$ of the change in conductivity of the NW on illumination to that in dark ($\delta = \Delta g_{NW}/R_o^{-1}$) is shown in Figure 3a for both the NW devices. Substantial change in the photoconductivity compared to that in dark is observed. Assuming that mobility does not change upon illumination, the ratio $\delta$ is the change in the carrier concentration on illumination ($\Delta n/n_0 \equiv (n - n_0)/n_0$), $n_0$ being the carrier concentration in dark. At high illumination intensity (> 20 μW/cm$^2$), $\Delta n$>$n_0$.

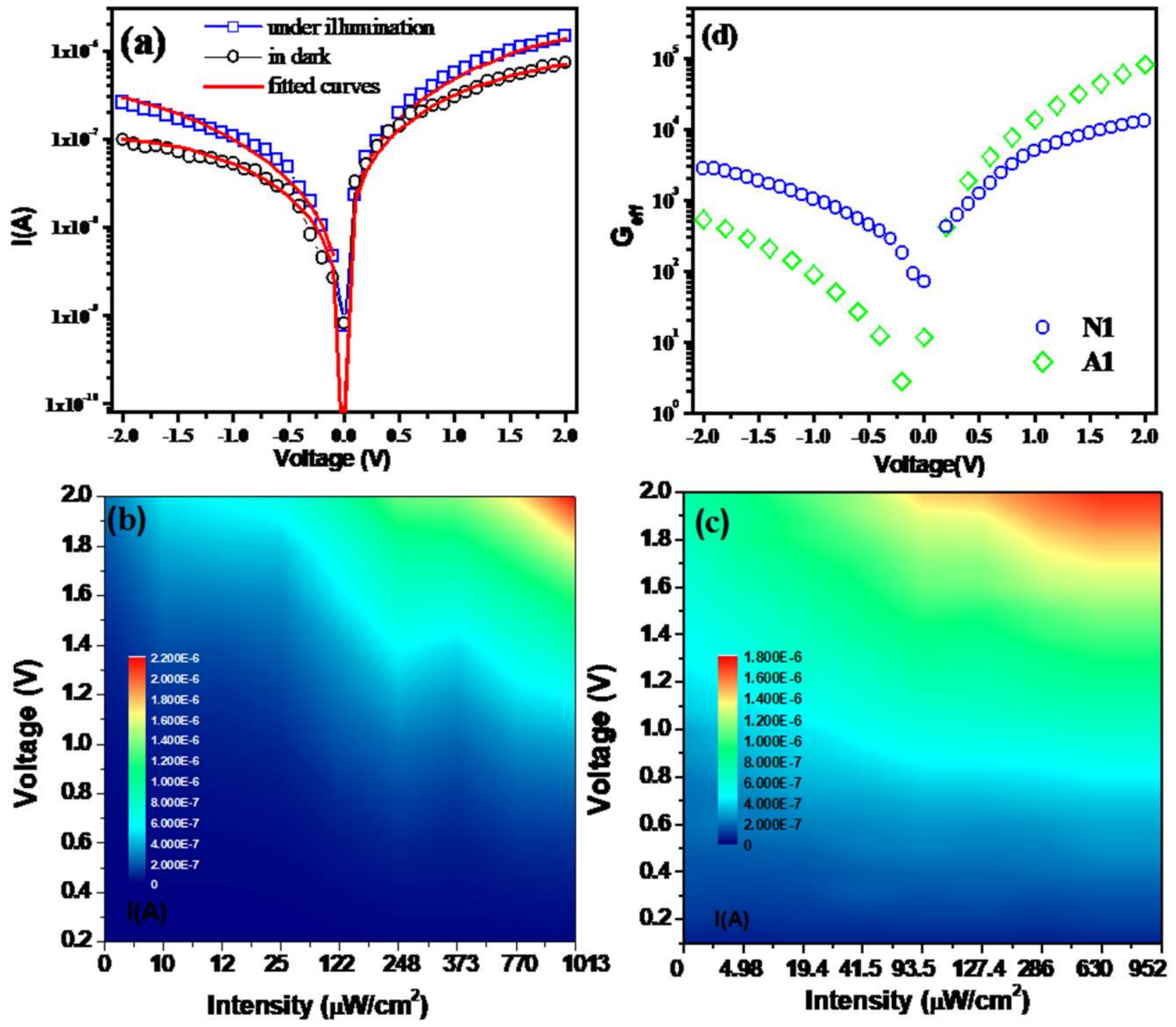

Figure 2. (a) ) I-V characteristics of device N1 in dark (black) and under illumination (blue) at 650nm wavelength of light with intensity ~ 100μW/cm$^2$ along with the MSM model fitted curves (red). (b) and (c) Contour plot of device current as a function of bias with light intensity for A1 and N1 respectively at 650nm wavelength of light (d) Effective Gain ($G_{eff}$) plotted as a function of bias voltage for both NWs at 650nm light for intensity ~ 100 μW/cm$^2$.

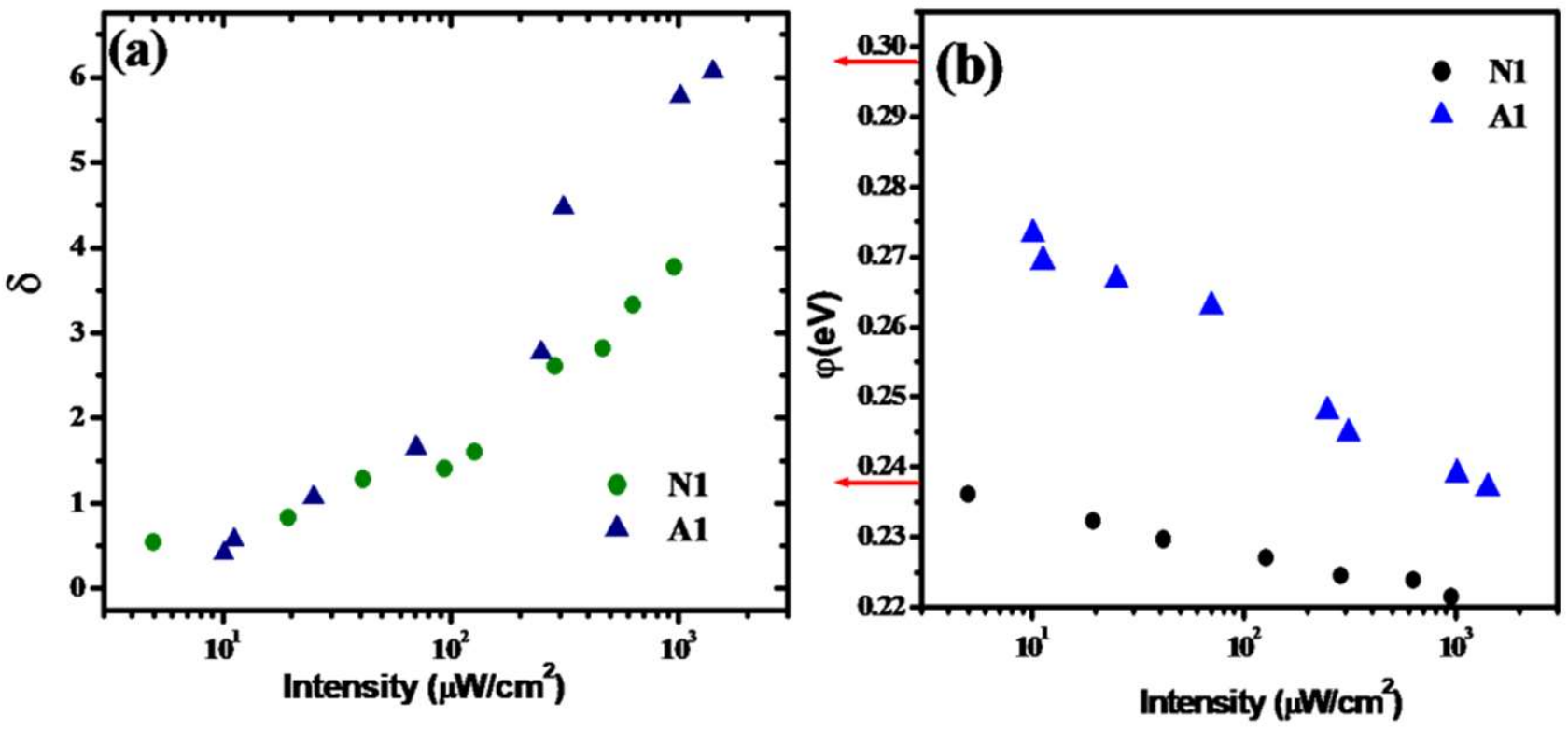

Figure 3. (a) The ratio δ (relative change in conductance on illumination) as a function of illumination intensity at 650 nm light. (b) Shows the barrier heights as deduced from analysis of the data as a function of illumination intensity. The barrier height at dark is marked with red arrow on y-axis.

In Figure 3b, we plot the Schottky Barrier (SB) height as obtained from the fit above, where it shows reduction of the SB as a function of illumination intensity. We have plotted the average barrier height $\phi \equiv (\phi_1 + \phi_2)/2$, where $\phi_1$and $\phi_2$ have been obtained from the modified Richardson-Dushmann equation (eqn. 1) fit to the data. The lowering of SB height occurs for both the nanowires. The initial barrier height, $\phi$ in dark for NW A1 is larger compared to that of the NW N1, although they are of comparable order. In case of NW N1, the lowering of the SB height $|\Delta\phi| \approx .02\ eV$and that for the NW A1 is $|\Delta\phi| \approx .06\ eV$for illumination intensity ~ 1mW/cm². To assess the reduction in the barrier height that occurs due to diffusion of carriers in the MS contact region, we estimate $|\Delta\phi|$, from change in the chemical potential due to the change in carrier density due to diffusion. The carrier diffusion upper limit is given by the change in the carrier concentration $\Delta n/n_0$ due to photo generated carrier which we estimated before. Thus, we estimate an upper limit of the SB height lowering due to the photo generated carriers as:[24]

$$|\Delta\phi| = \frac{k_B T}{q} ln\left(\frac{n_0+\Delta n}{n_0}\right) \quad (2)$$

The ratio$(n + n_0)/n_0)$ has been obtained from the experimentally measured $\delta$ as mentioned before. Comparison with observation shows that the values of $|\Delta\phi|$ though comparable to the estimated values, are consistently lower (~ 10meV) at 1mW/cm$^2$. This is expected; the calculated $|\Delta\phi|$ is an upper limit which assumes that all the photogenerated carriers diffuse into the contact region. In reality, a smaller fraction will diffuse in, because some part of the photogenerated carriers recombine in the depletion region and in traps. The comparison establishes that the diffusion of carriers in the MS contact region leads to changes in the chemical potentials which lower the barrier height. The reduction in $\phi$ is an important effect, which adds on to the photocurrent of the Ge NW photodetector.

### 3.3 Ultrahigh spectral response in UV-Vis-NIR region

The major result of the present investigation is ultra large responsivity $\mathcal{R}$ obtained in the photodetector made from a single Ge NW. A simple measure of the sensitivity of the single Ge NW can be appreciated from the following observation. A photocurrent ~ 50 nA is detected in a single Ge NW (A1) with area of light collection ~ $10^{-9}$ cm$^2$ at 2V bias in a room dimly lit with white light (from a CFL). The same photocurrent is obtained by a commercial Si photodetector with collection area of 1 mm$^2$. This immediately shows that the responsivity of the nanowire detector is larger than a commercial detector by few orders which is approximately the ratio of the collection areas of the Si detector and the Ge NW detector.

The Responsivity versus wavelength data is shown in Figure 4a for both the detectors. The data shown in Fig. 4a show that the NWs have photo response (R ~ $10^7$ A/W) that are orders of magnitude higher compared to those of photodetectors made from bulk Ge crystals (R<1 A/W ). The spectral response of both the NW extend to similar ranges although the value of $\mathcal{R}$ differ by two orders of magnitude. At a bias of 1V for NW N1, $\mathcal{R} > 10^4$ A/W over the broad spectral range of 300 nm to 950 nm peaking at a value of $\mathcal{R} = 10^5$ A/W at 850 nm. At the same bias (1V) and over the same spectral range, for NW A1, $\mathcal{R} > 10^6$ A/W, an order higher. This observation is in conformity with the $\mathcal{R}$ measured in different semiconductor single NW photodetectors which finds that decreasing the diameter enhances the value of $\mathcal{R}$. At 2V for NW A1, an enhancement of one more order has been achieved and $\mathcal{R} > 10^7$ A/W over the spectral range 300 nm to 1000 nm with a peak value of $\mathcal{R} = 3 \times 10^7$ A/W at 850 nm. We note that this value is one of the largest photo responsivity reported in single NW photodetectors that show broad spectral response. In the photodetector with diameter 30 nm NW, the value of $\mathcal{R}$ is at least an order larger compared to that of the detector with diameter 65 nm. The dip in $\mathcal{R}$ as shown in fig. 4a in the shaded region is due to source characteristics as we have a Xenon lamp source, which has very strong emission lines in the NIR region interposed with regions of low emission intensity. It is also noted that the decrease in photo response at longer wavelengths partly arise due to decrease in absorption in this region in Ge NWs.[16] To check the efficiency of the NW photodetector at longer wavelengths, we measured the photocurrent in the broad band NIR region from 980-2600 nm using a white light source. The $I_{Ph}$ obtained from broad band source in NIR region is ~1 μA as shown in Figure 4b for both the nanowires. This is much larger than the dark current and the noise limit. The broad band source used has peak emission at 1000nm from which we determine the intensity ~ 1.5 mW/cm$^2$, using Si detector (R= 0.46 A/W at 1000nm). The photocurrent of the order μA in both NWs, is quite large and can be assumed to have partly originated from illumination in the NIR region beyond 1100 nm and upto 1850 nm, which is the band gap of Ge. Photoconductivity measurements though performed at individual wavelengths in the NIR region (beyond 1100 nm), cannot be quantified due to lack of a proper intensity calibration.

There are certain factors which may lead to overestimation $\mathcal{R}$ of which, one is the physical cross-section of a laid wire as the light collection area because of the antenna effect due to the underlying Silicon (below the 300 nm $SiO_2$ layer).[25,26] We have performed a control experiment and shown that this effect does not hamper or interfere with photoconductive properties of the Ge NWs under present experimental conditions. We point out that in the device structures that have been used, there may be some additional contributions arising from the $SiO_2$ layer acting as a wave guide and thus collecting radiation over an effective area that is larger than the nanowire. To rule out these effects we have tested the photo-response by illumination away from the NW. We find that for illumination even at a distance of 2-3 times the diameter of the NW the response is very little. Thus, any uncertainty arising from such an effect is within a factor of 2-4. Thus, a responsivity of $3x10^7$ A/W can get scaled down to not less than $10^7$ A/W by such effect. Thus, the large photo-response is an intrinsic property of the Ge NW.

Enhancement of the photoconduction on illumination has also been measured through the photo-current gain as defined previously. An example of measured $G_{Pc}$ versus illumination intensity is shown in Figure 5a for the device with NW A1 taken at wavelength 650 nm measured at a bias of 1V. Similar data is also seen for the device N1. $G_{Pc}$ increases with bias as has been shown in supplementary data Figure S3. For a given bias (see Figure 5a) $G_{Pc}$initially increases with intensity, reaches a maximum and then reduces again at higher intensity. It should be noted that a part of the enhanced device current also arises due to lowering of SB height $\phi$ as discussed in the previous section. The contribution to the enhancement of device current due to enhancement of the barrier penetration probability can be measured by a factor, $\kappa = \exp(|\Delta\phi|/k_B T)$. Thus dividing the observed $G_{Pc}$ by this factor gives us the effective gain $G_{eff} \equiv G_{Pc}/\kappa$ . In Figure 2b, there is a plot of $G_{eff}$ as a function of bias voltage for both the nanowires. Figure 5a shows a plot of $G_{eff}$ as a function of the illumination intensity. At lower intensity $|\Delta\phi|$ being small, the difference is not much. However, at higher intensity where $|\Delta\phi|$ is comparable to $k_BT$, the correction is large. Nevertheless, we observe that the value of intensity where $G_{Pc}$ and $G_{eff}$ both show a peak remains unaltered.

In Figure 5b, we show the effective gains $G_{eff}$ for both the NW devices. $G_{eff}$ reaches a maximum value in excess of $10^7$. This is one of the highest reported till date and is surpassed only by single NW ZnO detector that reaches a peak gain of $10^8$.[5] In the device with NW A1 (diameter = 30 nm), $G_{eff}$ is larger by a factor of six as compared to that of the device with NW N1 (diameter=65 nm). However, the illumination intensity where, $G_{eff}$ peaks is very similar for both the wires at ~ 10 μW/cm$^2$.

The extremely large value of $G_{eff}$ ≥$10^7$ achievable in the photodetectors is contributed by a number of factors. The enhancement of carrier concentration on illumination ($\Delta n$) is definitely a large contributing factor, as has been discussed above. However, this alone cannot enhance the gain to such high values. Another important factor is contributed by the device size where the length $L$ (distance between electrodes) being ~1-3μm, is less than the recombination length that makes the transit time of the carrier $\tau_{tran} = L^2/\mu V$ short. A short $\tau_{trans}$ makes the external quantum efficiency ($EQE = \tau_{reco}/\tau_{tran}$, where $\tau_{reco}$is the carrier recombination time) large so that the gain $G_{eff}$ ($\propto EQE$) becomes large. The carrier diffusion length for both hole and electrons has been measured in intrinsic Ge NW[27] and for holes is ~4-5 μm, depending on the mobility of the material. As a result, devices with sizes in the range of 1 μm can collect carriers without sufficient recombination.

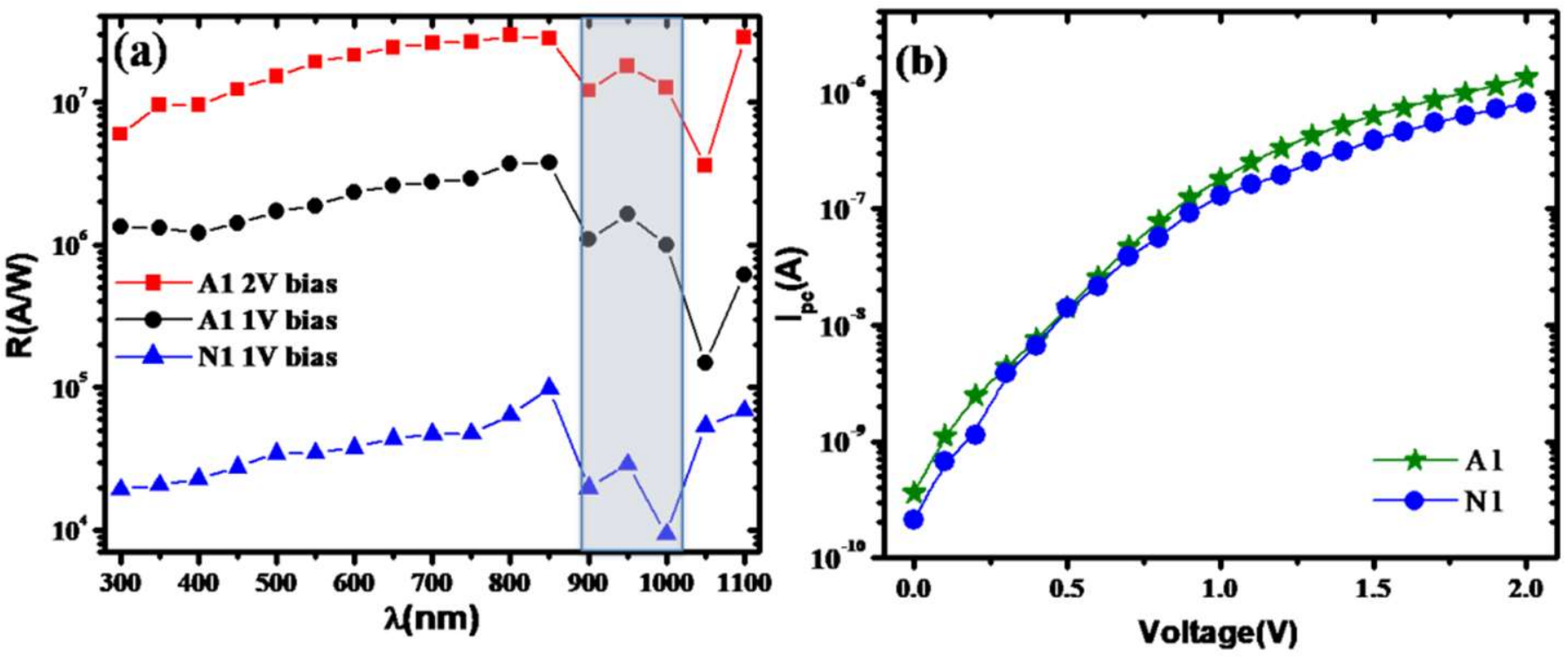

Figure 4. (a) Spectral dependence of Responsivity of the Nanowire photodetectors at different bias taken with illumination intensity ranging from 50 – 600 μW/cm². The shaded region has very low source intensity. (b) The photo-current at longer wavelengths in the NIR region measured with a broad band source (980nm -2600nm).

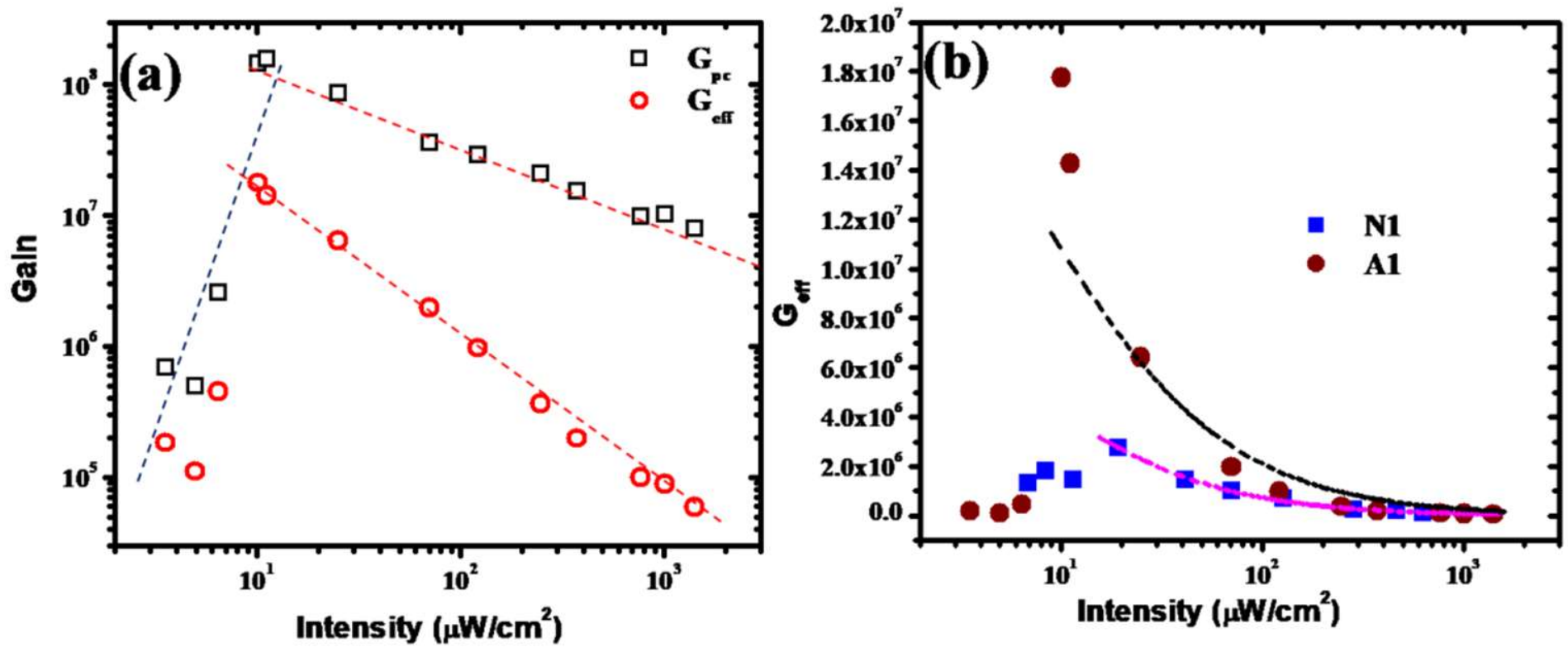

Figure 5. (a) Gain $G_{pc}$ and Effective Gain ($G_{eff}$) versus Intensity for the detector with NW A1 at a wavelength of 650nm for 1V bias. (Bias dependence of $G_{pc}$ is shown in supplementary data (see Figure S3). (b) Effective gain for the two nanowire devices as a function of illumination intensity. The dotted line is the fit to the equation 3 that gives decrease of effective gain due to saturation of interface states (see table 2).

### 3.4 Gain dependence on illumination intensity

Nanowire photodetectors differ from their bulk counterpart mainly due to the dominant role of surface. It is often taken as a recombination site and thus lowers photo response in bulk. However, in case of NWs, the surface plays a very different role and enhances the responsivity. The existence of a depletion layer on the surface of a NW (which is promoted by existence of a native oxide layer or one that is acquired during growth) plays an important role that is not seen in a bulk detector. It is the small radius that helps in using the surface states as a major facilitator[5] for enhancing the photoresponse. Presence of the depletion layer (and the special nature of the $GeO_x$/Ge interface) traps the electrons effectively and separates them from the photogenerated holes by an effective radial field (see Figure 6). This inhibits carrier recombination and the holes can reach the electrodes without significant recombination. While in a forward biased contact (+ve), the barrier for holes is high, in the reverse biased junction (-ve), the holes get effectively collected. This process of carrier separation by a built-in radial field due to formation of a depletion layer at the $GeO_x$/Ge interface, is at the core of enhanced performance of the NWs as an efficient photodetector.[5] As electrons get trapped at these interface states, the strength of the radial field decreases i.e. the depletion width decreases and this reduces the Gain at high intensities. This also explains why making the wire narrow (smaller diameter) improves its performance by a few orders. The existence of surface states acts as an effective facilitator for enhanced photoresponse and can be seen from the dependence of $G_{eff}$ on the illumination intensity where a peak in the gain is seen. For illumination intensity $\mathcal{J}$ > than the optimal value ($\mathcal{J}_0$), the gain decreases as a function of intensity. As the illumination intensity increases, the carrier generation rate increases which saturate the traps, reducing the band bending that gives rise to the radial field. As a result beyond a certain illumination intensity the gain decreases. The dependence of $G_{eff}$ with $\mathcal{J}$ for $\mathcal{J} > \mathcal{J}_0$ is given by,[5,12]

$$G_{eff} = \frac{\tau_l/\tau_{tran}}{1+\left(\frac{\mathcal{J}}{\mathcal{J}_0}\right)^{\alpha}} \quad (3)$$

where, the unsaturated intensity independent gain depends on the carrier transit time $\tau_{tran}$, $\alpha$ is a phenomenological fitting parameter that depends on the recombination of carriers at the trap states and $\tau_l$ is the carrier lifetime. The fitting of the experimental data to the above equation also gives us the value of the illumination intensity $\mathcal{J}_0$ where the gain saturates. From the value of $\mathcal{J}_0$ it is possible to obtain the value of surface trap density $T_d$ using the relation, $T_d = \mathcal{J}_o/h\nu$. The parameters obtained from the fits to equation 3 are given in table 2. For both the wires the values of $\mathcal{J}_0$ are very similar although it is somewhat larger for the device N1 which has relatively larger trap density. We do not have an independent measure for the trap density at the interface. The value of $\tau_l/\tau_{tran}$~ $10^7$, shows that the lifetime of carriers (holes) are indeed enhanced due to capture of electrons at the trap states, driven by the radial field. In the 30 nm NW, there is a tenfold increase in $\mathcal{R}$ at the same bias as compared to N1 which has 65 nm diameter. The above discussion clearly establishes the crucial role that a surface play in the photocarrier dynamics.

**Table 2. Fit parameters for dependence of effective on illumination intensity.**

| NW Name | $\tau_l/\tau_{tran}$ | $\mathcal{J}_0$(μW/cm$^2$) | $\alpha$ | $T_d$(cm$^{-2}$s$^{-1}$) |
|---|---|---|---|---|
| N1 | 1.0x10$^7$ | 9.0±2 | 0.87 | (2.9 ±0.2) x10$^{13}$ |
| A1 | 2.5x10$^7$ | 8.3±1.4 | 0.95 | (2.6 ± 0.2)x10$^{13}$ |

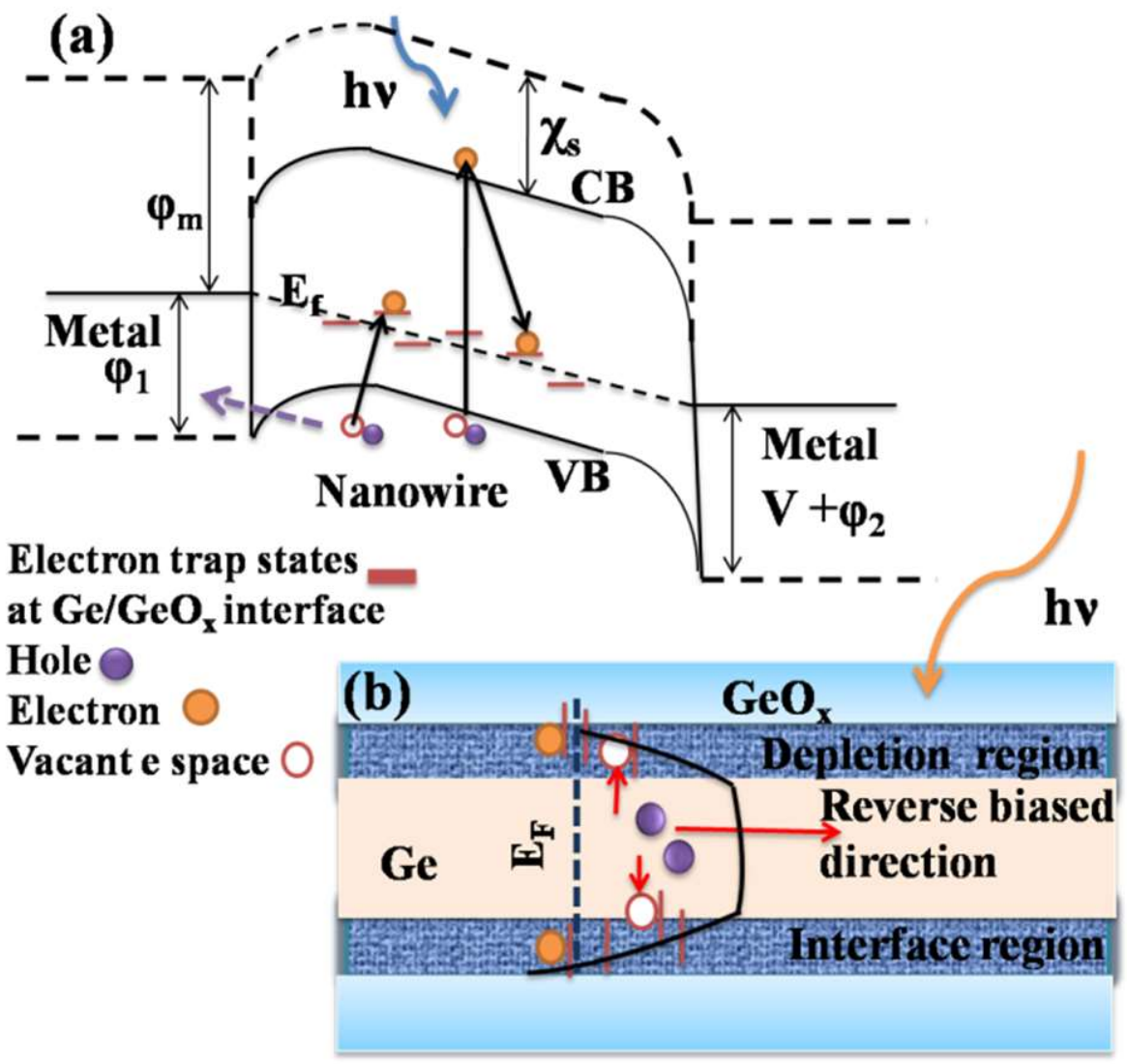

Figure 6. (a) A schematic of the axial section of the photodetector shows Schottky barrier at both ends of the metal/nanowire, with trap states lying near the Fermi level. The device is under illumination, showing dynamics of photogenerated electron and holes. (b) Photoconduction mechanism in Ge NWs showing trap states lying at the Ge/$GeO_x$ interface, which capture electrons. The radial field is created by a depletion layer at surface that drives electrons towards the traps, thereby separating the photogenerated pair, and migration of the hole to the reverse biased electrode. (The interface region of Ge/$GeO_x$ has been highly exaggerated in the schematic diagram.)

### 3.5 Nature of the oxide layer

From the nature of dependence of gain on illumination intensity, we proposed presence of an oxide layer on the surface of Ge NWs. To establish and substantiate the above hypothesis, we have performed Electron Energy Loss Spectroscopy (EELS) using TEM to get information on the chemical binding states of Ge and nature of the oxide layer. The EELS spectrum was analysed using multiple Gaussian peak fitting. The EELS spectrum of Ge L edge (figure 7a) shows a broad spectrum. The spectrum of Ge L edge corresponding to the transition of Ge 2p has been fit using standard Gaussian fitting, which shows Spin-orbit splitting into $2p_{3/2}$ ($L_3$) and $2p_{1/2}$ ($L_2$) at 1217 and 1247 eV respectively. Apart from these elemental peaks of Ge, there is another peak whose edge is 1222 eV and this corresponds to the binding energy of Ge other than metallic state.[28] The O K edge spectra correspond to the transition of O 1s is shown in Figure 7b. Splitting can be observed in the O K edge which shows two peaks, one of which is the K edge from O 1s orbital at 532 eV. The other peak shows absorption edge below 530 eV, which corresponds to binding energy of metal oxides. Thus it can be inferred from the EELS spectrum analysis that the Ge NW has oxygen, which is surface absorbed as well as in the form of $GeO_x$ ($x \leq 2$).

The spatial position of the Ge oxide was investigated using Energy dispersive X-ray Spectroscopy (EDAX) line profile of a Ge NW taken in Scanning TEM (STEM) mode along the radial direction, as shown red arrow in Figure 7c. The High-angle annular dark-field (HAADF) detector (Figure 7c) for imaging shows presence of a thin layer (~5nm) at a lower contrast compared to the bulk portion of the NW, which appears bright. In the inset of Figure 7c, the plot of The Line EDAX profile shows Oxygen concentration (from the K shell) at the two ends of the NW to be maximum, while the Ge concentration profile (from the L shell) is almost complementary to it. The maximum Ge concentration lies from 15-45 nm in the line EDAX profile. There is a finite size of the electron beam for each step size taken for the line profile which causes spatial broadening of the line scan leading to an apparent larger thickness of the oxide layer compared to the actual thickness. Nevertheless, it is established that the surface of the NW is surrounded by an oxide layer, while the core is Ge.

The efficiency of the surface states also becomes important due to the nature of $Ge/GeO_x$ interface. The $Ge/GeO_x$ interface is very different from the interface of $Si/SiO_2$. It was found through hybrid density functional theory calculations[26] that a majority of Ge and O atoms are threefold coordinated. This leads to formation of valence alternation pairs (VAPs) in the low-oxygen region of the interface.[29] These VAPs are prone to charge trapping; particularly it can trap a pair of electrons more efficiently than a single electron. This shows that in Ge NWs with an oxide surface, there is very efficient "scavenging" of electron from the generated pairs leaving the holes free to reach the electrodes. This drastically reduces carrier recombination and enhances photo-gain.

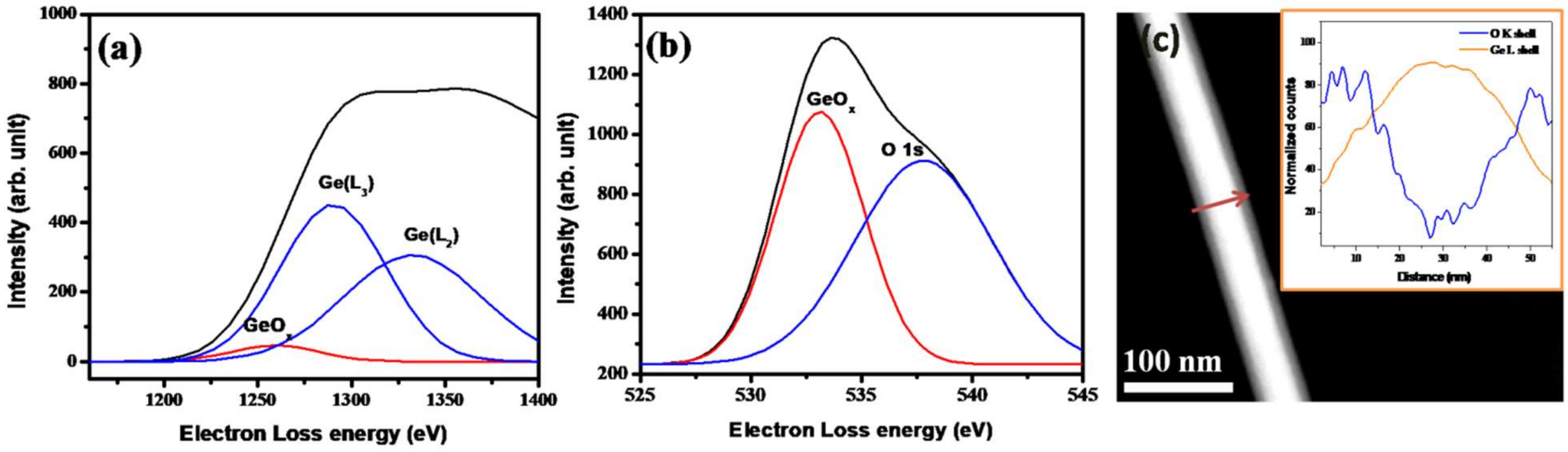

Figure 7. (a) EELS spectrum from 1150 to 1400 eV showing the Ge L3 (1247 eV) and L2 (1217 eV) peaks and a shallow oxide binding energy loss peak, with edge at 1220 eV. (b) EELS spectrum from 525-545 eV to capture the binding energy of Oxygen, which shows two peaks with an edge at 532 eV for adsorbed surface oxygen primarily from O 1s orbital and the other peak corresponding to a metal oxide peak, with edge below 530 eV. In both cases, the black line represents the experimental data, the blue lines are for elemental peaks and the red lines are showing other binding peaks. (c) The HAADF image of a Ge NW, showing the direction marked with red arrow, along which line EDAX profile was taken. Inset shows the line Edax of a single Ge NW taken in STEM mode with O and Ge concentration profile along the radial direction as marked by arrow.

### 3.6 Photon detection limit analysis

The effectiveness of the photodetector in presence of noise is an important factor, which is often measured by the Specific Detectivity, $D^* = \sqrt{\mathcal{A}\Delta f}/NEP$ as defined before. The signal band-width $\Delta f$ is selectable by the user and often is limited by the parasitic capacitance. In the device configurations used by us the measured $\Delta f$ is ≤ 1 KHz. We thus take an upper limit of the band width =1 KHz. As a realistic estimate we take *NEP* as the power at which the signal to noise ratio just reaches unity (see supplementary information Figure S4.). For the NW N1, *NEP* is 6x10$^{-15}$ W and for NW A1 device it is 0.3x10$^{-15}$ W. (All the relevant parameters are summarized in Table 1). The observed values of $D^*$ being larger than 10$^{11}$ cmHz$^{1/2}$/W and reaching even 10$^{13}$ cmHz$^{1/2}$/W for A1 is appreciable, since commercial bulk Ge detector has D ~10$^{12}$cm$^{1/2}$Hz/W.[11] This ensures that making the detector size in nanometre domain does not degrade its performance by extra noise. The value of $D^*$ is comparable to most single NW photodetectors.[4]

A practical guide to find the minimum energy (or minimum number of photons) the NW device can detect, may be obtained from the observed responsivity $\mathcal{R}$ = 3 ×10$^7$ A/W at 850 nm for NW A1 at an illumination intensity 57 μW/cm$^2$. For a detection band width of 0.5Hz (integration time ≈ 0.3 sec.) the rms current noise is ≈ 10nA. This implies that the detector can detect ≈ 0.3x10$^{-15}$ W and taking into account the integration time this will translate to 0.1x10$^{-15}$Joule. A photon of wavelength 850 nm carries energy 2.32 x 10$^{-19}$ Joule. Thus the detector can detect a minimum of ~430 photons. The highest gain $G_{eff}$ for the NWs occurs around 10μW/cm$^2$. For the detector with NW A1, the area $\mathcal{A}$ over which the light illumination is absorbed is ≈ 2.26x10$^{-9}$ cm$^2$. This translates to about 22x10$^{-15}$ W, which gives an upper limit for its best performance. Another alternate way to find the minimum power the detector is sensitive to, is to proceed from the measured gain $G_{eff}$. The detector retains a finite measureable gain even for illumination intensity of 1μW/cm$^2$. Thus it can detect below 2x10$^{-15}$W. These estimates give us a good range of the minimum energy/power the single Ge NW detector can measure and measurement of ~10$^{-15}$ Joule will be possible with these detectors, taking into consideration their *NEP* and detection band width.

### 3.7 Diameter dependent concentration the Electric field due to light inside a nanowire with cladding

In addition to the factors discussed before another additional factor contributes to enhanced photoresponse in narrow nanowires. There is partial enhancement of the electric field of the illuminating light thus increases electron-hole pair generation. We have used Maxwell's equations to find the analytical solution to the electric wave propagation inside a Ge NW with an oxide layer surrounding it. We have calculated the electric field in the cladding at 2.5 nm inside the $GeO_2$ layer and in the core at 3 nm from the central axis of the NW, which is 1μm in length. Details are given in Supplementary information.

From Figure 8, we can see that the amplitude of electric field is always greater in the core than the cladding. The electric field in the nanowire increases drastically as diameter reduces below 60 nm. In the 30nm wire the enhancement of the field is around 1.5 times than that in the 65 nm wire.

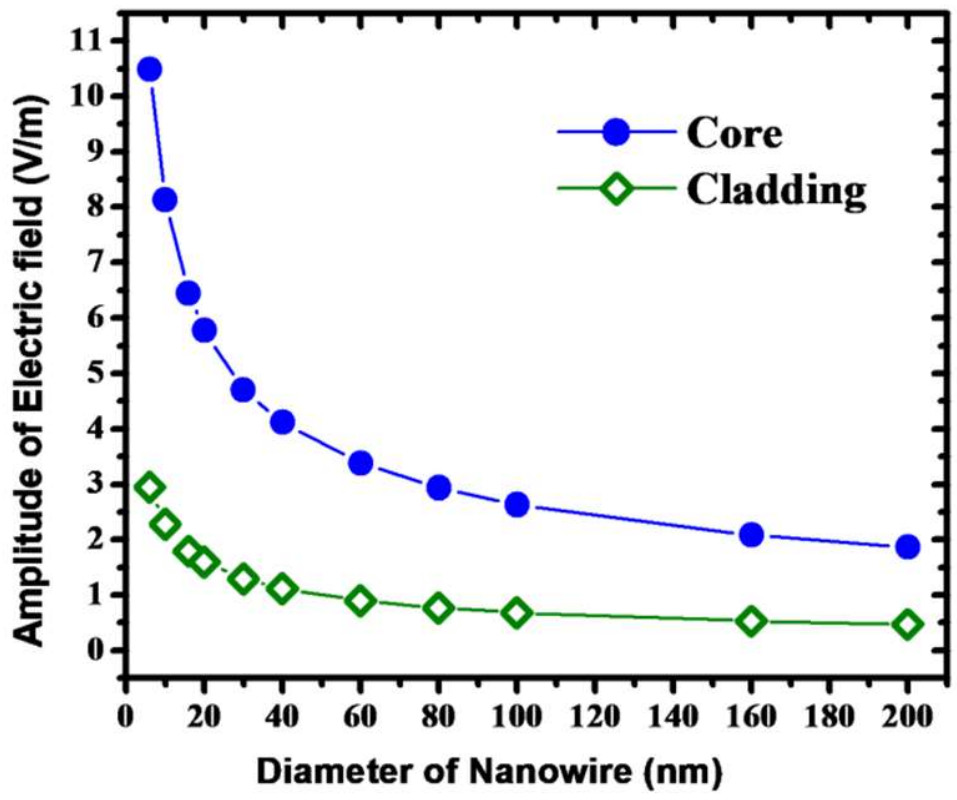

Figure 8. Amplitude of Electric field in the Ge NW core and the $GeO_2$ cladding layer

## 4. CONCLUSIONS

We establish that photodetectors can be made from single Ge NWs that show ultrahigh responsivity exceeding $10^7$ A/W (nanowire diameter ~ 30 nm) over a broad spectral range 300 nm -1100 nm with finite response extending well into the NIR region at wavelength > 1100 nm. The relatively low current noise and such high responsivity gives rise to a detector that can detect down to $10^{-15}$W. The large optical gain exceeding $10^7$ is achieved at a low optical illumination intensity ~ 10μW/cm$^2$. One of the core enabling factors that leads to such a high gain and responsivity is the surface layer of $GeO_x$ as we have seen from extensive electron microscopy analysis, which gives rise to a depletion layer that traps electrons of the illumination generated electron-hole pair and leaves the hole free to reach electrodes without recombination. The device length ~1-3μm being smaller than the carrier diffusion length facilitates collection of carriers effectively before recombination.

## SUPPORTING INFORMATION DESCRIPTION

The Supporting Information includes HAADF image of a nanowire, I-t curves of both nanowires at a fixed bias and different intensity, Gain as a function of intensity with varying bias and I-t plots for determination of NEP and details of the calculations in section 3.7. It is available free of charge on the ACS Publications website.

## ACKNOWLEDGEMENTS

The authors acknowledge financial support from Nanomission, Department of Science and Technology, Government of India, as a sponsored Project (SR/NM/NS-53/2010) and (SR/NM/NS-1141/2012(G)). AKR acknowledges additional financial support from Science and Engineering Research Board, Government of India as a J.C. Bose Fellowship (SR/S2/JCB-17/2006). The authors also acknowledge Centre For Energy Science (DST Nano mission project SR/NM/TP-13/2016). AG thankfully acknowledges SERB for funding as NPDF (PDF/2015/000179).

# Supplementary Information

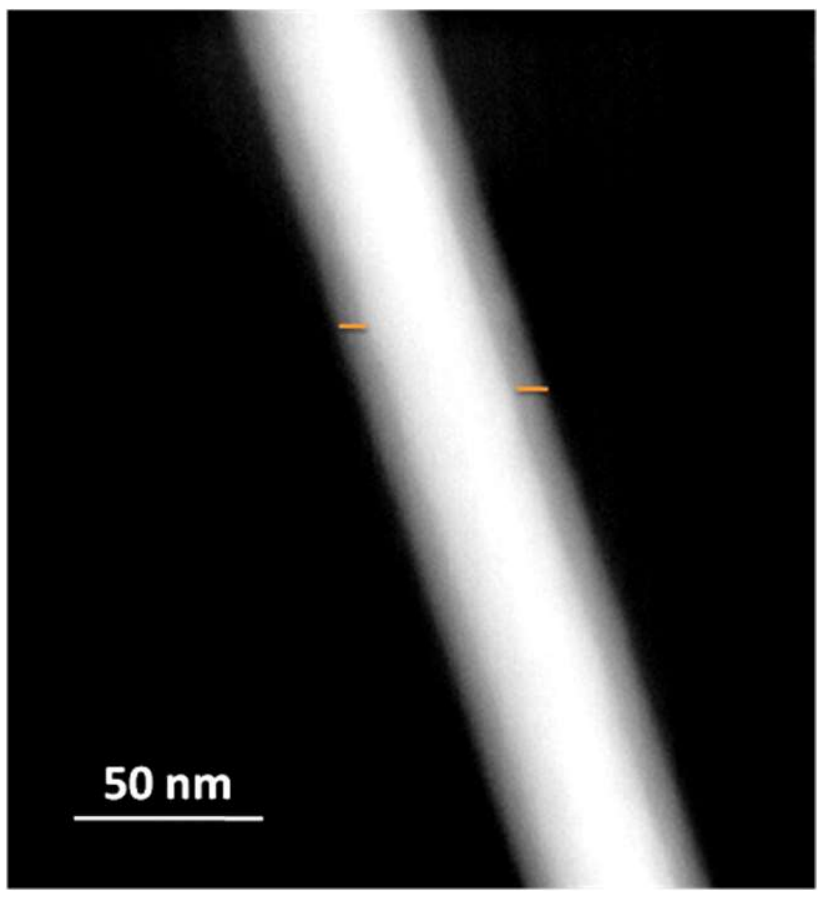

Figure S1. HAADF image here with the oxide layer marked in orange line. The oxide layer shows thickness of 6.5 nm while the core has thickness of 40 nm.

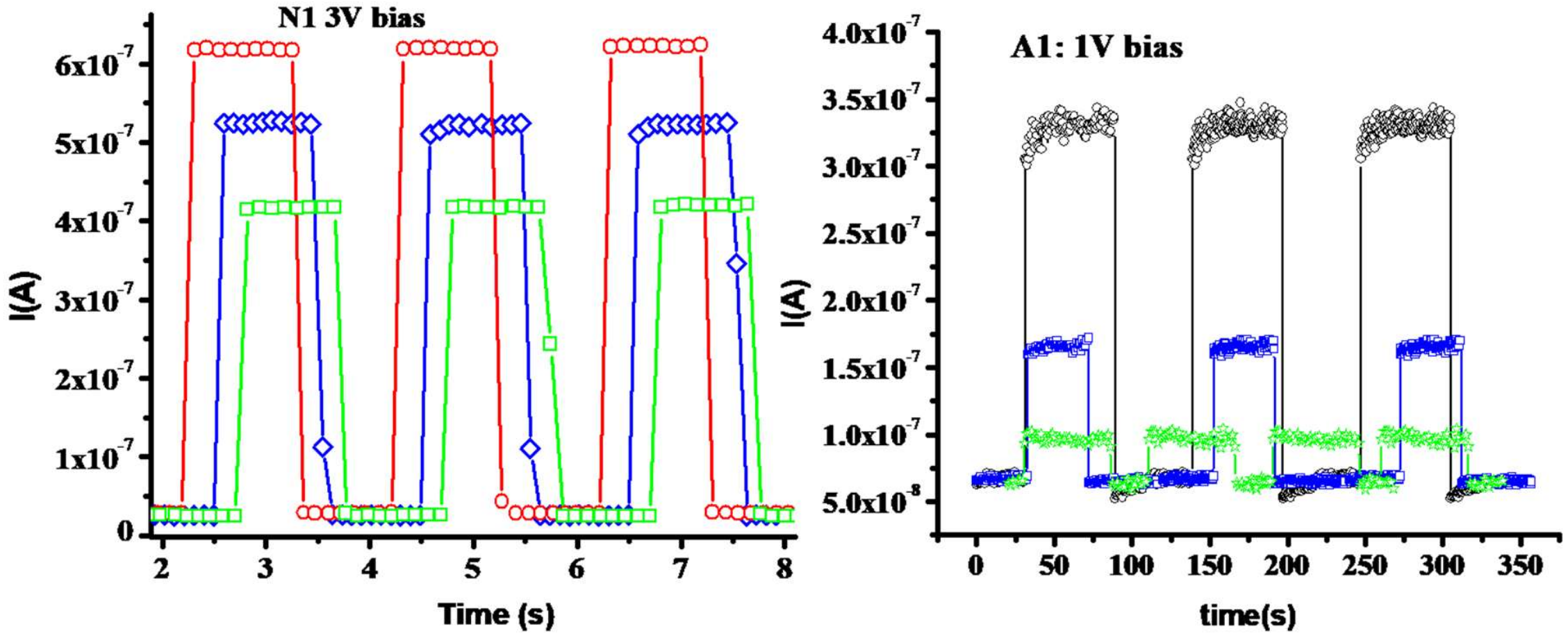

Figure S2. The I-t curves of N1 and A1 are shown for fixed bias at 650 nm light of different intensities.

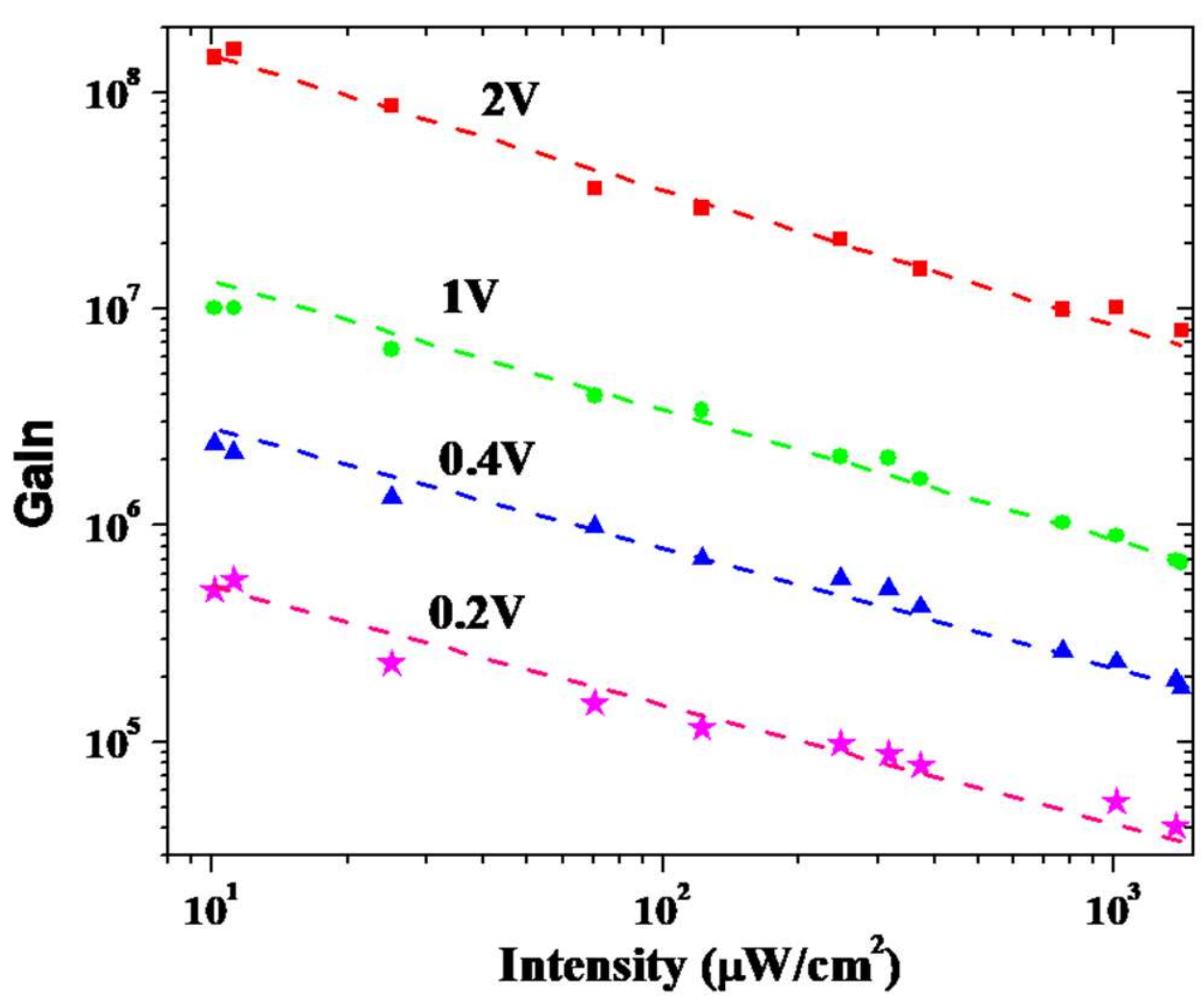

Figure S3. I-V characteristics of A1 at different intensities at 650nm wavelength of light.

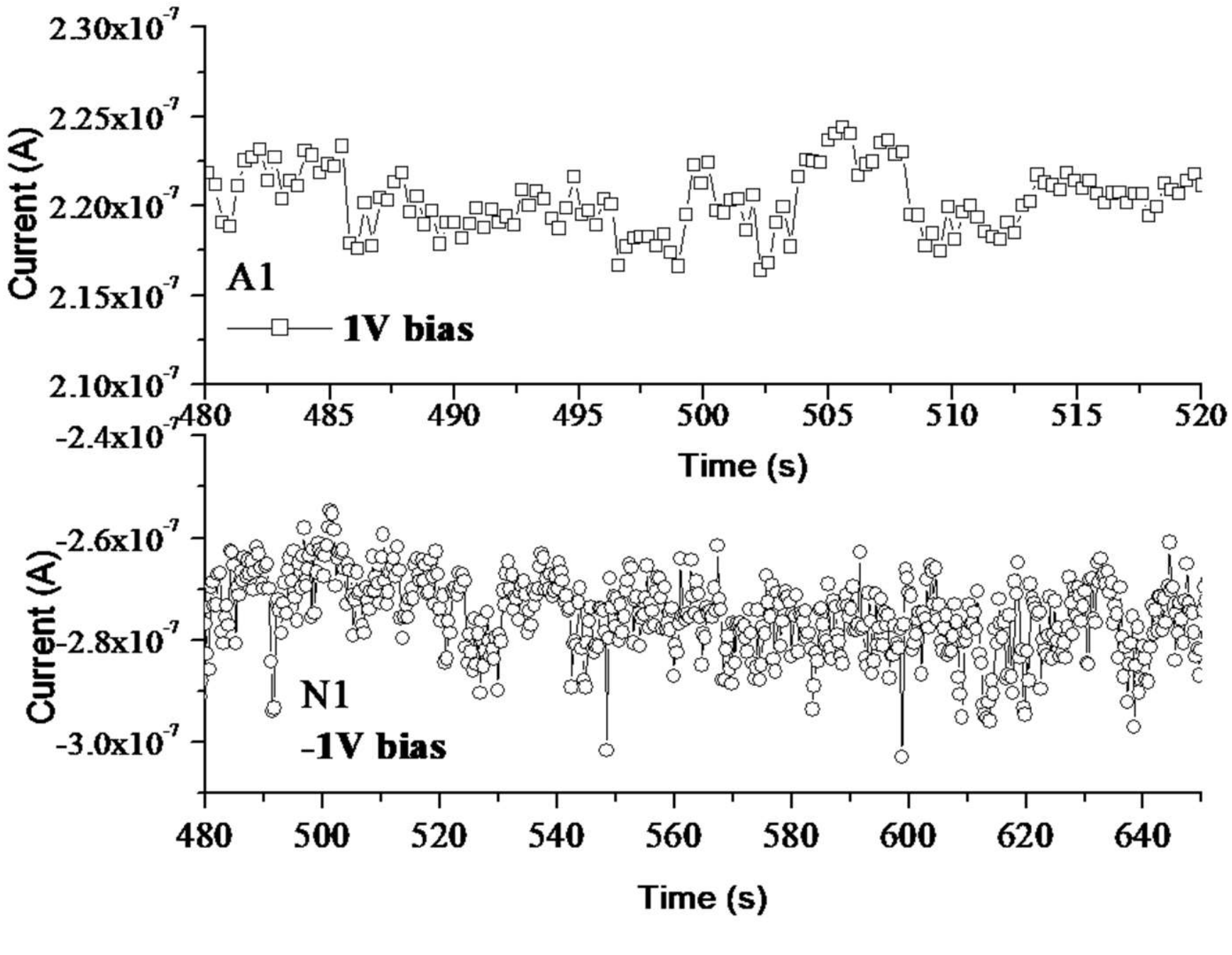

Figure. S4. The Current versus time plot from which we determine the NEP, when the signal to noise ratio is unity and the subsequent current noise in A1 and N1.

## *Supplementary note on:*

**Diameter dependent concentration the Electric field due to light inside a Nanowire with cladding.**

We have determined the amplitude of Electric field inside the NW at the core as well as the oxide as a function of nanowire diameter. We have taken a 2D model of rectangular slab of variable width (d), which is the Ge core diameter with 5 nm width of $GeO_2$ layer (cladding) on both sides of it. It takes into account an electromagnetic wave of wavelength 650 nm incident on the NW from one edge. The method assumes the electric wave propagation along the axis of the NW (z-direction) following the equation,

$$E_z = E_y e^{-ik_x x},$$

where, the y-component is along the radial direction. (see figure SS4 below). The NW has a surrounding oxide layer, which acts as cladding and causes a wave guiding effect in it. Hence, $E_y$ it is defined at the two regions as,

$E_y = Acos(k_y y)$ (in the Ge NW core) and

$E_y = Be^{-(\eta y+\frac{d}{2})}$ (in the $GeO_2$ cladding).

The continuity equation at the interface of the core and cladding gives us the wave guiding condition, solving which, we can determine η and in turn, the value of $E_z$. We have taken $3x10^4$ photons falling on unit area of the NW per unit time, to find the value of the constants A and B.

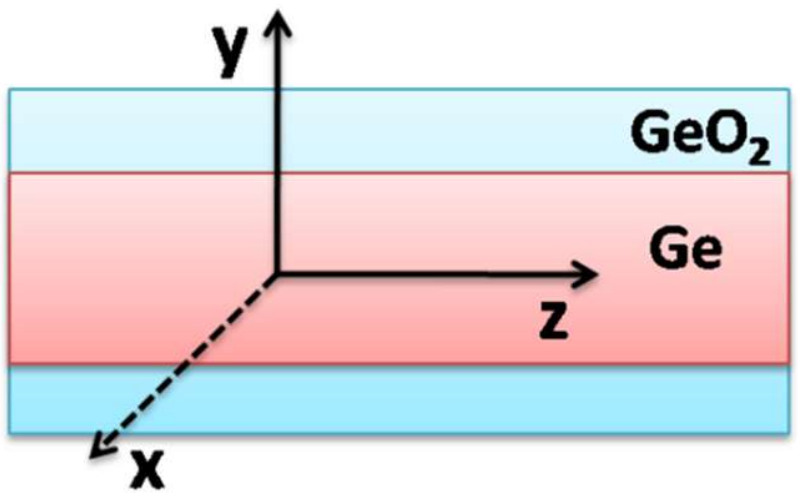

Figure S5. Schematic showing the coordinates in the Ge NW model which we used for calculations.